\documentclass[12pt]{article}
\usepackage[dvips]{graphicx}

\title{Spacetime and vacuum as seen from Moscow}
\author{L.B. Okun \\
ITEP, Moscow, 117218, Russia, email: okun@heron.itep.ru}
\date{}

\begin{document}
\maketitle

\begin{abstract}

An extended text of the talk given at the conference ``2001: A
Spacetime Odyssey'', to be published in the Proceedings of the
Inaugural Conference of the Michigan Center for Theoretical
Physics, University of Michigan, Ann Arbor, 21-25 May 2001, M.J.
Duff and J.T. Liu eds., World Scientific, Singapore, 2002; and of
Historical Lecture ``Vacuum as seen from Moscow'' at the CERN
Summer School, 10 August, 2001. Contents: Introduction;
Pomeranchuk on vacuum;
Landau on parity, P, and combined parity, CP;
Search and discovery of $K_L^0 \to \pi^+ \pi^-$;
``Mirror world";
Zeldovich and cosmological term;
QCD vacuum condensates;
Sakharov and baryonic asymmetry of the universe, BAU;
Kirzhnits and phase transitions;
Vacuum domain walls;
Monopoles, strings, instantons, and sphalerons;
False vacuum;
Inflation;
Brane and Bulk;
Acknowledgments; References.

\end{abstract}

\section{Introduction}

This talk presents: a few episodes from the history of ideas on
the structure and evolution of vacuum during last 50 years as seen
from ITEP; the impact on the theory of vacuum by I.Pomeranchuk,
L.Landau, Ya.Zeldovich, A.Sakharov, D. Kirzhnits and their
disciples; the importance of international exchange of ideas; the
crucial role of experiments on P and CP violation; interconnection
of particle physics and cosmology.

\section{Pomeranchuk on vacuum}

Being a student of Isaak Pomeranchuk I first heard from him in
1950: ``The vacuum is filled with the most profound physical
content" and ``The Book of Physics consists of two volumes: v.1.
Pumps and Manometers, v. 2. Quantum Field Theory.'' In the middle
of 1950's Pomeranchuk proved his famous theorem about equality of
particle and antiparticle cross-sections at the same target. It is
not an accident that the Regge pole with quantum numbers of
vacuum, which is responsible for this equality, was called later
Pomeranchuk pole or pomeron.

In 1956 Pomeranchuk encouraged Boris Ioffe, Alexei Rudik and
myself to work on P-non-conservation. By applying CPT theorem
\cite{1} to the famous paper of T.D. Lee and C.N. Yang \cite{2} we
came to the conclusions \cite{3} that: 1) $\vec{\sigma}\vec{p}$
correlations suggested by them would signal not only P-, but also
C-violation. 2) $K_S$ and $K_L$ are respectively even and odd
under T (or, better, CP), but not under C \cite{4}. Hence
$K_S\not\to 3\pi^0$, $K_L^0 \to 3\pi^0$ (November 1956). The same
conclusions were reached by T.D. Lee, R. Oehme and C.N. Yang
\cite{5} (January 1957) (see Nobel lectures by T.D. Lee and C.N.
Yang \cite{6,7} (December 1957)).

\section{Landau on parity, P, and combined parity, CP}

Till the end of 1956 Lev Landau did not believe in the possibility
of P-violation. In the summer of 1956 he did not encourage Iosif
Shapiro to publish a Moscow University report \cite{8}, in which
an experiment of the type carried out later by C.S.Wu et al. was
suggested (according to Shapiro's report energy of a system
changed its value under mirror reflection). Landau was convinced
that P must be conserved because vacuum is mirror symmetric. (One
would not expect this argument from an author of 1950 article with
Ginzburg \cite{9a} in which spontaneous symmetry breaking was
introduced into theory of superconductivity.)

We (IOR) discussed our paper with Landau. A few weeks after it was
submitted to JETP, Landau \cite{9} put forward (December 1956) the
idea of absolute CP-invariance and conservation of ``combined
parity". According to this idea the ``sin" of P-violation was
committed by particles and antiparticles (with opposite signs),
but the mirror symmetry of vacuum was preserved.

The idea was widely accepted. I considered it to be beautiful, but
on the other hand, I did not understand why the Lagrangian could
not have complex coefficients. Therefore in lectures given at ITEP
(1960-61) and Dubna (1961) \cite{10} and in the book \cite{11}
based on these lectures (1963) the importance of experimental
tests of CP was stressed, in particular search for $K_L \to 2\pi$.

\section{Search and discovery of $\mbox{\boldmath$K_L^0 \to
\pi^+ \pi^-$}$}

A special search at Dubna was carried out by E. Okonov and his
group. They have not found a single $K_L^0 \to \pi^+ \pi^-$ event
among 600 decays of $K_L^0$ into charged particles \cite{12}
(Anikina et al., JETP, 1962). At that stage the search was
terminated by administration of the Lab. The group was unlucky.

Approximately at the level 1/350 the effect was discovered by
J.Christensen, J.Cronin, V.Fitch and R.Turlay \cite{13} at
Brookhaven in 1964 in an experiment the main goal of which was
$K_L \to K_S$ regeneration in matter.

Thus absolute CP-invariance was falsified.

\section{``Mirror world"}

Still the appeal of Landau's idea of absolutely symmetric vacuum
was so strong that in 1965 Igor Kobzarev, Isaak Pomeranchuk and
myself suggested the hypothesis of a ``mirror world" \cite{14}. We
assumed CPA invariance, where $A$ [from ``Alice through the
Looking Glass"] transforms our part of the Lagrangian into its
mirror part.

Each of our particles has its mirror counterpart. The mirror
particles have between them the same electromagnetic, weak and
strong interactions as ours. In principle there might exist mirror
nuclei, atoms, molecules, stars, planets, galaxies, even mirror
life. Whether they actually exist depends on cosmological
evolution.

The possibility of the existence of both ``left protons'', $p_L$,
and ``right protons'', $p_R$, had been discussed by T.D. Lee and
C.N. Yang in two last paragraphs of their famous article \cite{2}.
But they believed that $p_L$ and $p_R$ can interact with the same
pion and the same photon. We have proved that this is impossible.

According to our original assumption, the only particles which
belong to both our and mirror worlds are gravitons. If there were
two gravitons, nothing would connect the two worlds and the idea
of a mirror world would have no physical consequences.

Why the graviton but not, say, a photon? As soon as you assume
that the photon is common to both worlds, you immediately come to
contradiction with experiments. Colliding $e^+ e^-$ through a
virtual photon would annihilate not only into our particles but
also into mirror ones. Besides, the electromagnetic radiative
corrections would be doubled destroying the beautiful agreement of
QED with experiment.

In 1982 I studied the oscillations between our photons and sterile
ones, ``paraphotons'' \cite{15'}; in 1983 -- considered a stronger
coupling between two worlds which could be caused by exchange of
some hypothetical neutral bosons \cite{15}. A number of authors
considered mirror particles as a substantial component of dark
matter. Many enthusiastic  papers about mirror matter were written
recently by R. Foot (see \cite{17'} and references therein)
without proper citation of ref. \cite{14}.

\section{Zeldovich and cosmological term}

Yakov Zeldovich was very excited when he realized that a
non-vanishing $\Lambda$-term is strongly required by quantum field
theory. In 1967 he estimated $\varepsilon_\Lambda$ on the basis of
pairs of virtual elementary particles in vacuum. First he started
with $\varepsilon_\Lambda \sim m^4_{Pl}$ and found enormous (124
orders) discrepancy with observations (private communication).
Then \cite{16} he assumed that quartically divergent loops ($\sim
m^4_{Pl}$ and even $m^4$) vanish and estimated $$
\varepsilon_\Lambda \sim \frac{m^6}{m_{Pl}^2} \;\; , $$ where $m$
is the mass of a typical elementary particle of the order of a
proton mass (1 GeV). That was the first attempt to put $m_{Pl}^2$
in denominator. (The contribution of the loop is proportional to
gravitational interaction.) In order to bridge the gap of $10^8$,
he tried $$ \varepsilon_\Lambda \sim GG_F m^8 \;\; , $$ where $G$
is Newton constant ($m_{Pl} = G^{-1/2}$), while $G_F$ is Fermi
constant.

In 1968 Zeldovich noticed the negative sign of the contribution of
fermions and indicated that the contributions of bosons and
fermions might cancel each other! \cite{17} This was before Yuri
Golfand and his student Evgeni Likhtman published \cite{18'} the
first paper on supersymmetry.

Recently Johannes Bl\"{u}mlein and Paul S\"{o}ding kindly informed
me about references \cite{18, 19}, in which early unpublished
estimates made by young W. Pauli are described. Pauli discussed
vacuum fluctuations and their influence on cosmology. Note that he
overlooked that cosmological term would lead to inflation and
concluded that the universe would be smaller than the distance
from the earth to the moon. Even earlier, in 1916, the ``zero
point energy'' had been considered by W. Nernst (\cite{20}, see
also \cite{21}).

\section{QCD vacuum condensates}

Here I will make a digression from cosmology to Quantum
Chromodynamics (QCD) with its quarks and gluons.

In 1965 V. Vanyashin and M. Terentyev \cite{van} discovered the
negative sign of the vacuum polarization of gauge bosons in the
case of SU(2) gauge symmetry. Their numerical result was slightly
incorrect, as the ``ghost'' was unknown at that time. A correct
number (with the same negative sign!) was derived by I.
Khriplovich in 1969 by using Coulomb gauge \cite{khrip}. The
negative sign of vacuum polarization was a novelty: for photons in
QED it is positive. But at that time the nonabelian theories with
gauge vector bosons were not considered seriously at ITEP. Even
after SLAC experiments on deep inelastic scattering and their
parton interpretation by J. Bjorken \cite{27a} and R. Feynman
\cite{27b} we were not clever enough to realize the importance of
negative vacuum polarization.

The situation has drastically changed in 1973 when H. Fritzsch, M.
Gell-Mann, and H. Leutwyler \cite{27c} suggested QCD with its
SU(3) color gauge symmetry. In  papers by H.D. Politzer \cite{pol}
and D.J. Gross and F. Wilczek \cite{gross} the negative sign was
rediscovered, and its importance for QCD was stressed: it lead to
decreasing color charge at short distances (and hence to
asymptotic freedom at high momentum transfers).

It should be noted that Vanyashin and Terentyev \cite{van}
considered as ``absurd'' not the ultraviolet, but the infrared
behaviour of the charge where its square first rises and then
becomes negative. At present we know that this region is never
reached: in Electroweak Model because of the Higgs mechanism, in
QCD because of confinement. The increase of charge at large
distances makes perturbation theory invalid at such distances.
Non-perturbative effects manifest themselves in non-vanishing
vacuum expectation values (VEVs) of colorless products of quark
and gluon fields, the so-called QCD condensates. These condensates
were used at ITEP by Michael Shifman, Arkady Vainshtein and
Valentin Zakharov in order to successfully describe the properties
of various hadrons \cite{shif1}. This work was followed by many
hundreds of theoretical papers (for a review see \cite{shif2}).

\section{Sakharov and baryonic asymmetry of the universe, BAU}

In 1967 I had the privilege to witness the creation of Sakharov's
seminal paper \cite{22} on the Baryonic Asymmetry of the Universe
(BAU). He started out from the CP-violating charge asymmetry of
the semileptonic decays of $K_L^0$ and from the nonequality of
branching ratios in the decays of $\Sigma^+$ and
$\widetilde{\Sigma^+}$ discussed by S.Okubo \cite{23}. Using it as
a springboard, he jumped from strange particles to the universe
and from strangeness violation to baryon number violation. A small
excess of baryons over antibaryons in the hot big bang soup
survived, while the bulk annihilated into photons, neutrinos and
antineutrinos. Photons are observed today as CMBR. The cosmic
background neutrinos are tantalizing. When devising the baryon
nonconserving ad hoc interaction Sakharov was helped by ITEP team.
In the framework of SM and its extensions  there are several
mechanisms of baryon number violation; among them grand
unification \cite{24} and triangle anomalies \cite{25}. The latter
mechanism raises serious problems \cite{26} at electroweak scale:
it might wipe out baryon asymmetry, unless the original lepton
asymmetry is involved \cite{35a}. The problem of BAU can not be
solved in the framework of Standard Model. (V. Rubakov and M.
Shaposhnikov \cite{43}.)

\section{Kirzhnits and phase transitions}

With the advent of the Standard Model based on $SU(3)\times
SU(2)\times U(1)$ gauge symmetry new concepts came to particle
physics; one of them is the vacuum expectation value (VEV) of the
Higgs field. It reduces $SU(2)_L \times U(1)_Y$ to $U(1)_{em}$ and
provides with masses leptons, quarks, $W$ and $Z$ bosons. An
important step was made \cite{27} in 1972 by David Kirzhnits, a
theorist at the Lebedev Institute, who realized that the Higgs VEV
is a function of temperature and has to vanish in the early
universe, when we go backwards in time to temperatures much higher
than the present value of VEV. Hence the masses of particles have
also to vanish. This idea was further developed by Kirzhnits and
his student Andrei Linde \cite{28} and later by many others
\cite{29}. (Both A. Sakharov and D. Kirzhnits were disciples of
Igor Tamm.) I first heard about the vacuum phase transition from
David on a street during a conference in Tashkent in 1972.

\section{Vacuum domain walls}

\begin{figure}
\begin{center}
\includegraphics[width=0.5\textwidth]{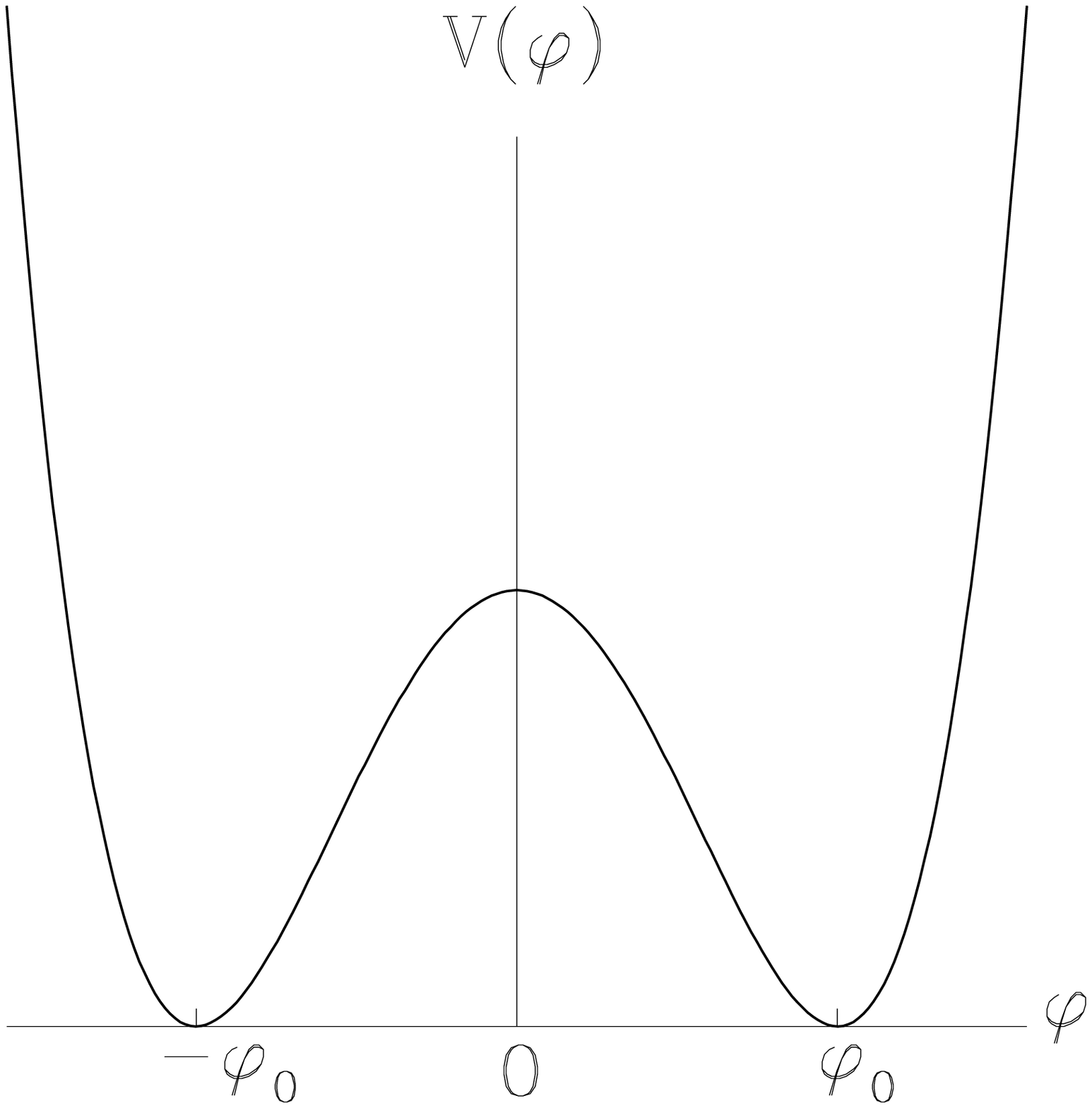}
\caption{\label{fig1}}
\end{center}
\end{figure}

Two years later Igor Kobzarev, Yakov Zeldovich and myself
\cite{30} merged the idea of vacuum phase transition with a model
of spontaneous CP violation proposed by T.D.Lee \cite{31}.
According to this model a neutral pseudoscalar field has two
degenerate vacua (see Fig. \ref{fig1}).

The vacuum around us has a sign that was fixed by chance during
the cooling of the universe and formation of VEVs. But at any
distant enough place the other sign might have been chosen. The
domains of different sign are separated from each other by
``domain walls" the thickness and density of which are determined
by VEVs of the field $\varphi$ and by its self-coupling. Typical
electroweak scale is several hundred GeV. We discussed the
evolution of the universe filled with domain walls and concluded
that at present the nearest wall should have gone beyond the
horizon, leaving as a farewell an anisotropy of CMBR. The advent
of inflational cosmology (see below) greatly weakened this
argument against spontaneous CP-violation.

\section{Monopoles, strings, instantons, and \\
 sphalerons}

Vacuum domain walls were the first megascopic cosmological
elementary objects considered in the framework of quantum field
theory. They were caused by spontaneous breaking of discrete
symmetry. Later in 1974 G.'t Hooft \cite{32} and A.Polyakov
\cite{33} discovered magnetic monopoles as solutions of
spontaneously broken SU(2) gauge symmetry. Cosmological creation
of monopoles and of strings (the latter appear as solutions of a
broken U(1) gauge symmetry) was considered in 1976 by T.W.B.Kibble
\cite{34}. (Microscopic U(1) strings of Nambu type underlying dual
resonance model of hadrons were considered in 1973, without any
reference to cosmology, by H.B. Nielsen and P. Olesen \cite{44a},
who were inspired by Abrikosov's vertices in a superconductor
\cite{44b}.)

Since the early 1980's a whole ``industry of theoretical papers''
on cosmological role of domain walls, strings and monopoles
appeared. I will mention here only articles by Ya.B. Zeldovich
\cite{35}, A. Vilenkin \cite{36} and an exotic example of a cosmic
string invented by Albert Schwartz \cite{37} in 1982 after he
learned from me about the ``mirror world". A particle, after
making a circle around his ``Alice string", transforms into a
mirror particle and becomes invisible for a terrestrial observer
\cite{38}.

A special type of non-perturbative transitions between
topologically different vacua of non-abelian gauge fields were
constructed by A. Belavin, A. Polyakov, A. Schwartz, V. Tyupkin
\cite{39} in 1975 and by 'Hooft \cite{40} in 1976. The latter
called them instantons. In the four-dimensional eucledian space
(with imaginary time) instantons represent solutions with finite
action. Many attempts to understand the mechanism of confinement
in QCD have exploited instantons.

Amplitudes of high temperature transitions between vacua in
electroweak theory, called sphalerons, were first analyzed by N.
Manton \cite{41} in 1983 and by F. Klinkhamer and N. Manton
\cite{42} in 1984 in connection with baryon nonconserving phase
transition at electroweak scale.

\section{False vacuum}

\begin{figure}
\begin{center}
\includegraphics[width=0.5\textwidth]{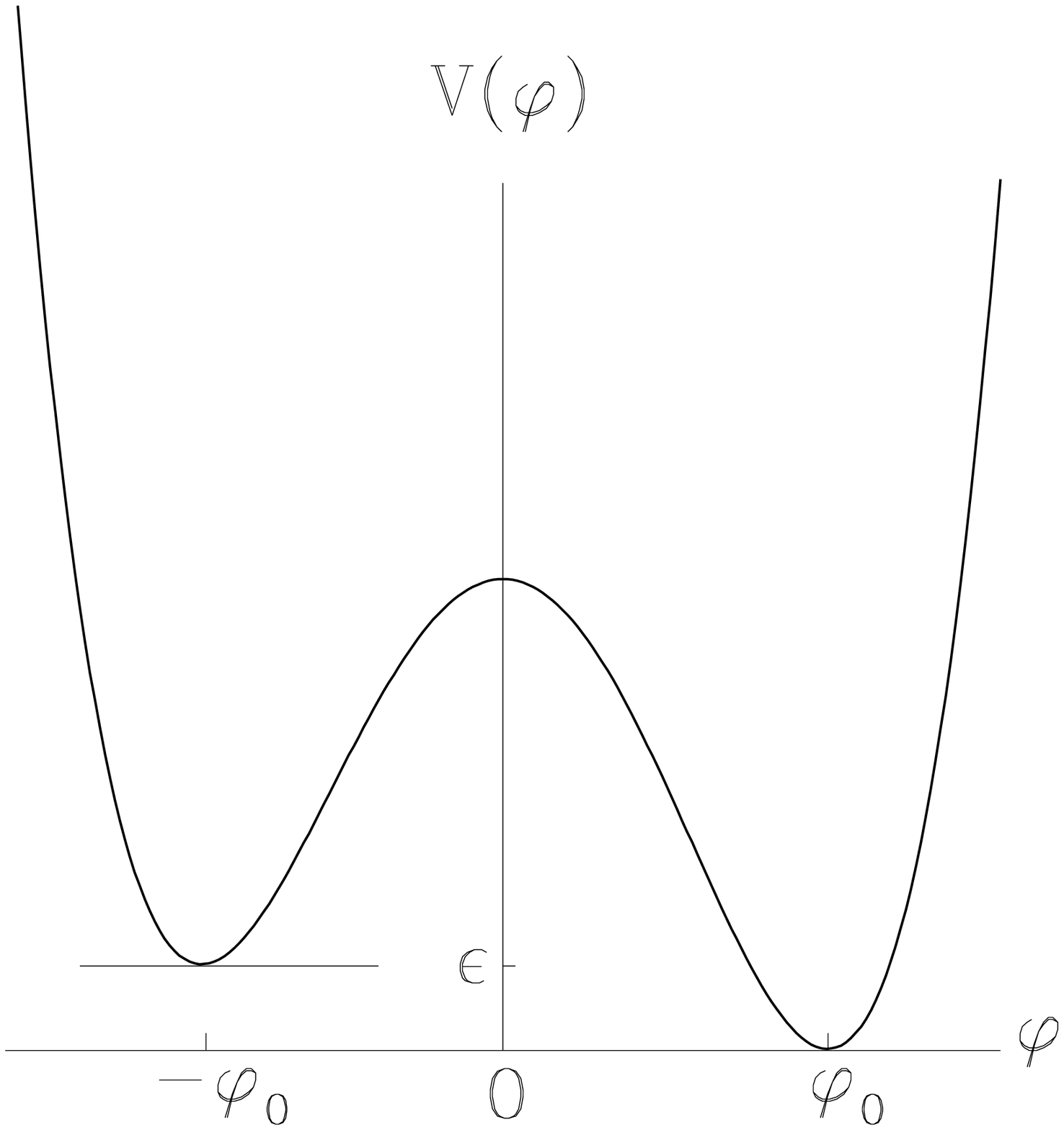}
\caption{\label{fig2}}
\end{center}
\end{figure}

In 1974 T.D.Lee and G.C.Wick \cite{44} suggested a Higgs potential
with two vacua, one of which lies a little bit higher than the
other (see Fig. \ref{fig2}).

The cosmological consequences for universe that lives in an upper
metastable vacuum were elaborated \cite{45} by Igor Kobzarev,
Mikhail Voloshin and myself in 1974 and later, in 1977, by Curt
Callan and Sidney Coleman \cite{46, 47} (they dubbed it ``false
vacuum") and also by Paul Frampton \cite{48}.

The spontaneous decay of a false vacuum starts by formation
through quantum tunneling of the smallest bubble of the true
vacuum surrounded by a wall which separates the two vacua. The
critical radius of this bubble is such that the gain of energy
proportional to its volume becomes large enough to compensate the
mass of the wall which is proportional to its surface: $$
\frac{4\pi}{3}R_c^3 \cdot \varepsilon = 4\pi R_c^2 \sigma \;\; ,
$$ where $\varepsilon$ is the energy density of the false vacuum
(see Fig. 2), while $\sigma$ is the surface density of the wall.
After that the bubble expands classically, destroying the
universe.

When I first thought that the creation of a bubble could be
catalyzed at a collider, my back shivered. Then I reassured
myself: all possible collisions have already occurred in the early
universe. A few months later I told Andrei Sakharov about the
bubble. His reaction was: ``Such theoretical work should be
forbidden". My argument about collisions in the early universe was
rejected by him: ``Nobody had collided two nuclei of lead".

In 1984 P.Hut \cite{49} published an estimate of uranium-uranium
collisions in cosmic rays during the time of the existence of
universe. It turned out that the number of such collisions, $N$,
was many orders of magnitude larger than at any future heavy ion
collider: $N \sim 10^{47}$ for Fe with $E\geq 100$ GeV/nucleon.
Experiments at RHIC (BNL) study Au - Au collisions at $\sqrt{s}
\sim 200 A$ GeV $\sim 40$ TeV ($A\simeq 197$ for gold). There will
be $2 \cdot 10^{11}$ collisions in 10 years, first at 30 GeV, then
at 70 GeV, and finally at 100 GeV/nucleon. A fixed target NA 50
experiment at CERN gathered $2\cdot 10^{12}$ Pb - Pb collisions at
$\sqrt{s} \sim 17 A$ GeV$\sim 3.5$ TeV ($A\simeq 207$ for lead).
The experiment ALICE at LHC (CERN) will study Pb - Pb collisions
at $\sqrt{s} \sim 1200$ TeV. The abundances of heavy nuclei may be
$10^{-5} - 10^{-10}$ of that of Fe, which still leaves a huge
safety factor in cosmic rays. For a recent update of these
arguments see \cite{50, 51}.

\section{Inflation}

The notion of a false vacuum with large energy density (VEV)
naturally led Alan Guth \cite{52} to the idea of inflationary
universe (1980). The exponential expansion solves the problems of
flatness and the horizon as well as homogeneity and isotropy
problems. It also explains the vanishing abundance of magnetic
monopoles. (For earlier models of exponential expansion of the
universe see papers by E.B. Gliner \cite{62a} and A.A. Starobinsky
\cite{62b}.)

In 1982 Andrei Linde \cite{53} and independently Andreas Albrecht
and Paul Steinhardt \cite{54} suggested how to stop this expansion
by producing many vacuum bubbles and high entropy. Also in 1982
Alexander Dolgov \cite{55} suggested a mechanism with a rolling
scalar field which led to what later was dubbed ``quintessence''.

Recent measurements with high angular resolution of CMBR
fluctuations confirm the inflation scenario (experiments BOOMERANG
\cite{56}, MAXIMA \cite{65a}, DAST \cite{65b}). The flatness of
the universe is corroborated by observations of high-z supernova
of type Ia \cite{57, 58}. On the other hand, they give a puzzling
result: the fractional energy density of the universe $\Omega =
\varepsilon/\varepsilon_c$, where $\varepsilon_c$ is the critical
density, corresponding to the flat universe, is dominated by
cosmological term -- i.e. energy density of vacuum ($\sim$ 70\%),
then follows the so-called dark matter ($\sim$ 30\%), while
ordinary baryonic matter constitutes only about 3\%. To understand
the cause and mechanism of such fine tuning is a great challenge.

During the 1990s the development of inflation cosmology has lead
to a new view on the place of our universe in the world. According
to this view, our universe is a part of eternal metauniverse,
which consists of innumerable universes. They create their
offsprings by forming Big-Bang bubbles in false vacua. The
properties of vacua, fields, particles and even number of
space-time dimensions differ from one universe to the other
\cite{60a}. This enormous  variety may explain the anthropic
properties of our universe which is so nicely tuned for our
existence (see, e.g. \cite{60b}).

\section{Brane and bulk}

The idea of extra dimensions \cite{61', 62'} has been actively
discussed for 60 years before it was first combined with the idea
of ``domain walls'' in these extra dimensions \cite{59},
\cite{60}. According to this idea, our four-dimensional world is a
``brane'' in a multidimensional bulk \cite{61} - \cite{63}. All
known particles and their fields are confined (completely, or
almost completely) within the brane, while real and virtual
gravitons are allowed to leave the brane and to propagate in the
bulk. This drastically changes gravity. The Planck scale may
become as low as 1 TeV. There could be deviations from the
standard Newton's force at distances of the order of millimeters
and many other strange phenomena. But at present all this is a
kind of a daydreaming.

\section*{Acknowledgments}

I am grateful for their hospitality to Mike Duff who invited me to
the Inaugural Conference of the MCTP and to Guido Altarelli and
Marcella Diemoz who invited me to give the Historical Lecture at
the CERN Summer School.

\end{document}